# Dimensionality-suppressed chemical doping in 2D semiconductors: the cases of phosphorene, MoS$_2$, and ReS$_2$ from first-principles

Ji-Hui Yang and Boris. I. Yakobson[*]

Department of Materials Science and Nanoengineering, Rice University, Houston, Texas 77005, USA

E-mail: biy@rice.edu

**Abstract:** In spite of great appeal of two-dimensional (2D) semiconductors for electronics and optoelectronics, to achieve required charge carrier concentrations by means of chemical doping remains a challenge, due to large defect ionization energies (IEs). Here by decomposing the defect IEs into the neutral single-electron defect level, the structural relaxation energy gain, and the electronic relaxation energy cost, we propose a conceptual picture that the large defect IEs are caused by two effects of reduced dimensionality. While the quantum confinement effect (QCE) makes the neutral single-electron point defect levels deep, the reduced screening effect (RE) leads to high energy cost for the electronic relaxation. The first-principles calculations for monolayer, few-layer, and bulk black phosphorus (BP), MoS$_2$, and ReS$_2$ with strong, medium, and weak interlayer interactions, respectively, as examples, do demonstrate the general trend. Based on the gained insight into defect behaviors, strategies can be envisaged for reducing defect IEs and improving charge carrier doping. Using BP monolayer either embedded into dielectric continuum or encapsulated between two $h$─BN layers, as practical examples, we demonstrate the feasibility of increasing the screening to reduce the defect IEs and boost carrier concentration. Our analysis is expected to help achieving effective carrier doping and thus to open ways towards more extensive applications of 2D semiconductors.

## INTRODUCTION

Two-dimensional (2D) semiconductors have emerged as promising novel functional materials for various applications (*1—4*). They have been experimentally demonstrated useful for key applications such as field effect transistors (FET) (*5—8*), catalysts (*9, 10*) and sensors (*11—13*). However, 2D semiconductors are still far behind three-dimensional (3D) semiconductors in many fields, i.e., solar cells, thermoelectrics, light-emitting diodes, transparent conducting materials, etc. Even for electronic applications, the prospects of 2D semiconductors for high-performance transistors are still controversially debated (*14, 15*). One of the main reasons which hinder 2D semiconductors from outperforming 3D "peers" is the challenge to effectively dope 2D semiconductors to n-type or p-type via chemical methods, which lies in the heart of semiconductor technologies, especially for junction-based applications (*16*) with large-scale integrations and reliable performances.

To overcome the above challenge, the defect properties of 2D semiconductors must be understood because defects provide free carriers and thus life to a semiconductor (*17*) while some defects behave as carrier killers (*18, 19*) and scattering centers. One of the most important defect properties is the defect IE, also known as defect transition energy level, defining the ability of a defect to be ionized to provide carriers (*17, 20, 21*). Defects with small IEs are needed to introduce sufficient carriers.

From 3D to 2D, reduced dimensionality can have great effect on the properties of semiconductors. For example, it can result in markedly enhanced electron–electron interactions, which have been demonstrated to generate giant bandgap renormalization and excitonic effects from both many-body theoretical calculations and experiments (*22—25*). Similarly, defect IEs in 2D semiconductors can also be strongly affected by quantum confinement effect (QCE) and the reduced screening effect (SE) due to the reduced dimensionality. As Fig. 1 illustrates, when the dimensionality is reduced from 3D to 2D, the delocalized valence band maximum (VBM) and conduction band minimum (CBM) tend to shift downward and upward, respectively, due to the increased QCE. Meanwhile, the defect state, which can be either strongly or weakly localized within material layers (*26*), is less affected by QCE compared to the band edges. As a result, the energy difference between the defect state and the band edges, which contributes significantly to the defect IE, will always get larger due to the QCE. At the same time, the reduced dimensionality can result in the decrease of the dielectric constant and reduce the screening interaction between the ionized defect and all the other charges. According to the hydrogen model (*27*), the IE of shallow defects can be approximately written as $\frac{m^* q^4}{8\epsilon_s^2 \epsilon_0^2 h^2}$, where $m^*$ is electron effective mass, q is defect charge amount, $\epsilon_s$ is dielectric constant, $\epsilon_0$ is vacuum permittivity, and h is Planck's constant, respectively. With the decrease of dimensionality and thus the decrease of the dielectric constant (while the material is replaced by vacuum), the defect IE is expected to become larger even without QCE. The overall effect due to dimension reduction leads to the much larger defect IEs in 2D semiconductors, making the doping for charge carriers difficult. Note that, depending on to what extent the defect states are affected by QCE and SE, the defect transition energy levels can go either upward or downward in the bandgap with the dimensionality reduction, as shown in Fig. 1.

To overcome the charge carrier doping difficulties caused by low dimensionality, it is important to demonstrate through factual computations and understand the trends shown in Fig. 1. Here, by taking mono-, few-layer and bulk BP, $MoS_2$, and $ReS_2$ as prototype systems and using first-principles calculation methods, we explore the above concepts and reveal how QCE and SE affect the defect IEs. As typical examples of the most promising post-graphene 2D materials, BP, $MoS_2$, and $ReS_2$ have attracted intensive research and shown promising properties for various applications (*5—13, 28—51*), many of which require that they must be doped with a high carrier concentration especially for optoelectronics (*31—35, 41,48—51*). The three systems with strong, medium, and weak interlayer interactions in BP, $MoS_2$, and $ReS_2$, respectively, as demonstrated by their large, medium, and small band edge position changes from bulk to monolayers (Fig. S1), can serve as ideal systems to study the defect behaviors when dimensionality is reduced with the decrease of layer numbers.

To obtain the defect IE, we use the common supercell approach for defect calculations to get the total energy $E(\alpha, q)$ for a supercell containing the relaxed defect $\alpha$ in its charge state $q$. For faster convergence of total energies and defect IEs and good description of the symmetry of the defect states, we adopt a mixed scheme (*20, 52*). In this scheme, for an electron acceptor ($q < 0$), the defect transition energy level with respect to the VBM is given by

$$\varepsilon(0/q) = [\varepsilon_D^\Gamma(0) - \varepsilon_{VBM}^\Gamma(host)] + \{E(\alpha, q) - [E(\alpha, 0) - q\varepsilon_D^k(0)]\}/(-q), \quad (1)$$

where $\varepsilon_k^D(0)$ and $\varepsilon_D^\Gamma(0)$ are the defect level at the special $k$ points (averaged) and at the $\Gamma$ point, respectively; $\varepsilon_{VBM}^\Gamma(host)$ are the VBM of the pristine BP supercell at the $\Gamma$ point. The first term on the right-hand side of Eq. (1) gives the neutral single-electron defect level $E_{neu}$ at the $\Gamma$ point. The second term determines the $U$ energy parameter of the charged defect calculated at the special $k$ points, which is the extra energy cost after moving $(-q)$ charge to the neutral defect level with $E = \varepsilon_k^D(0)$. It can be further decomposed into the Coulomb or electron relaxation contribution and structural relaxation contribution, i.e., $U = [E(\alpha, q) - E^0(\alpha, q)] + \{E^0(\alpha, q) - [E(\alpha, 0) - q\varepsilon_D^k(0)]\}$, where $E^0(\alpha, q)$ is the total energy of the charged defect supercell when its atomic positions are the same as those in the fully relaxed neutral defect supercell. By this decomposition, the first term mainly includes structural relaxation contribution $E_{SR}$, and the second term mainly includes electron relaxation contribution $E_{ER}$. The defect IE can thus be written as $\varepsilon(0/q) = E_{neu} + E_{SR} + E_{ER}$. Note that, $E_{neu}$ can be largely affected by QCE while $E_{ER}$ is strongly related to SE. On the other hand, $E_{SR}$ is less affected by the reduced dimensionality. For a donor ($q > 0$), the defect transition energy level referenced to the CBM is given by

$$\varepsilon_g^\Gamma(host) - \varepsilon(0/q) = [\varepsilon_{CBM}^\Gamma(host) - \varepsilon_D^\Gamma(0)] + \{E(\alpha, q) - [E(\alpha, 0) - q\varepsilon_k^D(0)]\}/q, \quad (2)$$

where $\varepsilon_{CBM}^\Gamma(host)$ is the CBM of the pristine BP supercell at the $\Gamma$ point. The donor IE can be similarly decomposed into the above three contributions.

**RESULTS**

Without loss of generality, we mainly use neighboring-element-substituting point defects as our study targets: for BP, we consider extrinsic acceptor Si substitution of phosphorus ($Si_P$) and extrinsic donor Te substitution of phosphorous ($Te_P$) (we didn't use S because experimental result of $Te_P$ in bulk BP is available for comparisons with our calculations). Besides, we also consider the intrinsic defect phosphorus vacancy ($V_P$); for $MoS_2$, we consider the acceptor $P_S$ and the donors $Cl_S$ and $Re_{Mo}$; for $ReS_2$, we only consider extrinsic acceptor P substitution of sulfur ($P_S$) and extrinsic donor Cl substitution of sulfur ($Cl_S$) for simplicity due to the large supercell sizes. For the case of $V_P$ in BP, our previous works (*53, 54*) show that the most stable structure is the reconstructed $V_P$ as shown in Fig. 2(a), which is in agreement with recent results by Hu *et al.* (*55*). In this case, three dangling bonds are left when a P atom is removed. Then two P atoms get closer to each other and form a new P─P bond, leaving only one P dangling bond and creating a defect level near the VBM, which can accept one electron from the VBM, thus acting as an acceptor (*54*). The structures of $Si_P$ of $Te_P$ are also shown in Fig. 2(a). In bulk BP, the calculated (0/-) level of $V_P$ is 0.02 eV above the VBM and the calculated (0/+) level of $Te_P$ is 0.04 eV below the CBM. Our results are in good agreement with the experiment in which the intrinsic acceptor activation energy is 0.018 eV and the Te doped BP has donor activation energy of 0.039 eV (*56*). The calculated (0/-) level of $Si_P$ is 0.09 eV below the VBM. The very small defect IEs in bulk BP are expected from its small bandgap, which is only 0.30 eV. For bulk $MoS_2$, The calculated (0/-) level of $P_S$ and (0/+) level of $Cl_S$ are 0.10 eV above the VBM and 0.16 eV below the CBM, respectively. The (0/+) level of $Re_{Mo}$ is 0.11 eV below the CBM, in good agreement with other works (*57*). These levels are relatively shallow compared to the calculated bandgap of bulk $MoS_2$, which is 1.11 eV compared to the experimental value of 1.29 eV (*49*). For $ReS_2$, inequivalent sulfur sites are considered for the substituting defects $P_S$ and $Cl_S$ and the most energetically stable sites, as shown in Fig. 3(a), are used to study their defect transition energy levels. The calculated (0/-) level of $P_S$ and (0/+) level of $Cl_S$ in bulk $ReS_2$ are 0.15 eV above the VBM and 0.14 eV below the CBM, respectively, which, again, are relatively shallow compared to the bulk bandgap of 1.29 eV (experimental value is around 1.35 eV (*43*)).

To directly track the dimensionality effect on the defect behaviors, it's necessary to look into how defect properties change with different layer thickness: from monolayer (true 2D with n = 1), to bi-, tri- (n = 2, 3) etc. layers as a transition, and to bulk with n = ∞, that is true 3D. Here we limit the number of layers within three due to the very large supercell sizes required for defect calculations, i.e., the trilayer BP supercell contains about 300 atoms (see Supplemental Materials). Moreover, in few-layer systems, defects can be created in different layers: for example, in Figs. 2(b) and 3(b), defects can be created at three inequivalent sites, labeled as surface, *in*, and *in'*. At different sites, defects experience different QCEs and SEs and thus have different defect IEs, as well as $E_{neu}$, $E_{SR}$, and $E_{ER}$. In the following, we will investigate how these terms change with different layer thickness and different defect positions.

First, we present the results of BP systems which always have direct bandgaps from bulk to monolayer. Because the band edges of BP are mainly constituent of $p_z$ orbitals of P atoms which have large distributions out of BP layers (Fig. S2a), the changes of both VBM and CBM positions are very large from bulk to monolayers, indicating the strong interlayer interactions. As shown in Fig. 2(c), when BP changes from bulk to monolayer gradually, the defect transition energy levels with respect to band edges become deeper and deeper. For example, for the defects at the surface sites or in bulk BP, the (0/-) levels of $^{surf}V_P$ ($^{surf}Si_P$) increases from 0.02 (-0.09) eV in bulk BP to 0.10 (0.16) eV in trilayer, to 0.19 (0.30) eV in bilayer and to 0.57 (0.68) eV in monolayer BP. The (0/+) level of $^{surf}Te_P$ increases from 0.04 eV in bulk BP to 0.60 eV in trilayer, to 0.69 eV in bilayer and to 0.75 eV in monolayer BP. Similar trends are also observed for the defects at the *in* sites. Note that, the relative large defect IEs suggest that it's difficult to chemically dope monolayer BP to have a high carrier concentration. For example, the hole density can be calculated from $p_0 = N_v e^{-\frac{E_F}{k_B T}}$, where, the effective density of states of the valence band that can accept holes, is defined as $N_v = \int_{-\infty}^{0} d\varepsilon \left[1 + e^{(-\varepsilon)/k_B T}\right]^{-1} D(\varepsilon)$ with $D(\varepsilon)$ being density of states. In intrinsic BP with $V_P$ as the dominant defect that can be activated, the maximum hole density at T = 300 K can only reach $2.01 \times 10^3\ cm^{-2}$ (or $3.77 \times 10^{10}\ cm^{-3}$ using the BP monolayer thickness of 5.33 Å), assuming the Fermi level is pinned at $V_P$ (0/-) level with calculated $N_v$ of $6.67 \times 10^{12}\ cm^{-2}$.

In $MoS_2$, the bandgaps experience indirect-to-direct transitions from bulk to monolayers (*49*). The VBM position is always located at Γ point except that it shifts to the K point in the monolayer (*58*). However, the VBM energy difference between Γ and K points in the monolayer is very small (*58*). As a result, we reference all the acceptor levels to the VBM at Γ point for consistence and comparison. Similarly, the CBM position slightly moves from an intermediate point along the Γ—K line in the bulk to K point in the monolayer with very small energy difference (*58*). Consequently, we reference all the donor levels to the CBM at K point. The VBM at Γ point (different from K point) not only has $d_{z^2}$ orbitals of Mo atoms which are limited within MoS$_2$ layers, but also is largely constituent of out-of-plane $p_z$ orbitals of S atoms which has large distribution out of MoS$_2$ layers, as seen in Fig. S2(b). As a result, the VBM position is largely pushed downward from bulk to monolayer due to interlayer interactions. On the other hand, the CBM at K point only has $d_{z^2}$ orbitals of Mo atoms limited within MoS$_2$ layers (Fig. S2b). Therefore, its position is much less sensitive to the layer number changes with slightly upward-shifting from bulk to few-layers. In monolayer MoS$_2$, the CBM is slightly pushed downward due to the slight lattice expansion (If we fix the lattice constants, the monolayer CBM also shifts slightly upward from bulk to monolayer, as shown in Fig. S1). The large and small changes of VBM and CBM positions, respectively, indicate the medium interlayer interactions in MoS$_2$. But still, the defect transition energy levels with respect to band edges also become deeper with reduced layer numbers in MoS$_2$, as shown in Fig. 3(c).

In ReS$_2$, the bandgaps are kept (nearly) direct with both the VBM and CBM positions located at Γ point from bulk to monolayer. Both the VBM and CBM states are mainly constituent of Re *d* orbitals which only have distributions within ReS$_2$ layers. As a result, the changes of both the VBM and CBM positions are small from bulk to monolayer, indicating weak interlayer interactions in agreement with experimental reports (*43*). Despite this, the defect transition energy levels with respect to band edges still become deeper with reduced layer numbers and the defect IEs are very large in monolayer ReS$_2$, as shown in Fig. 4(c) (Note that, only results of surface defects are shown for simplicity).

The reason why defect IEs get larger with the reduction of dimensionality is twofold. Firstly, with the decrease of the layer thickness, the bandgaps increase due to the QCE along $z$-direction. The defect states, depending on the distribution localization within or out of material layers, can be affected by QCE as well. Because the acceptor (donor) states are generally derived from the valence (conduction) band, they are expected to follow the change of VBM (CBM) (*59*). As seen in Fig. 2(d), the absolute values of the $E_{neu}$ for $^{surf}V_P$ and $^{surf}Si_P$ in BP systems become smaller, following the reduction of VBM states when the layer thickness is reduced. What's different is that, the absolute value of the $E_{neu}$ for $^{surf}Te_P$ shows less change compared to the upshift of the CBM states when the BP thickness is reduced from trilayer to bilayer and to monolayer. This is because $Te_P$ has a defect state much more localized within BP layers than $V_P$ and $Si_P$, as clearly seen in the Fig. 2(a) and thus QCE has the weakest effect on the $E_{neu}$ of $Te_P$. We noice that the $E_{neu}$ of $Te_P$ even move downward when BP changes from the bulk case to trilayer. To understand this, we find that the conduction-band-derived defect states of $Te_P$ arise from the coupling of Te $p_z$ orbitals and P $p_z$ orbitals through the vertical Te—P bond (see Fig. 2a). From bulk BP to trilayer BP, the vertical Te—P bond length is largely increased from 2.623 Å to 3.033 Å. As a result, the coupling between the $Te_P$ and VBM state gets much weaker, leading to the reduced $E_{neu}$ of $Te_P$. When BP gradually changes from trilayer to monolayer, the Te—P bond lengths show little changes within 0.020 Å. Together with the fact that $Te_P$ defect state is strongly localized within BP layers, the $E_{neu}$ of $Te_P$ shows little changes in few-layer BP. As comparisons, the vertical Si—P bond lengths for valence-band-derived $Si_P$ defects have very small changes within 0.040 Å when BP changes from bulk to few-layer and monolayer. As a result, the coupling between the Si $p_z$ orbitals and P $p_z$ orbitals through the vertical Si—P bond is unlikely to change the $E_{neu}$ of $Si_P$ but the QCE dominants.

In $MoS_2$, we find that the QCE is very weak for the donor defects and the CBM states because all these states mainly have distributions localized within $MoS_2$ layers, as seen in Fig. S2(b) and Fig. 3(a). As a results, the $E_{neu}$'s of both $Cl_S$ and $Re_{Mo}$ remain nearly unchanged with the changes of $MoS_2$ layer numbers. In the contrast, the acceptor $P_S$ has a large defect state distribution out of $MoS_2$ layers (Fig. 3a). Consequently, the $E_{neu}$ of $P_S$ follows the downshift of the VBM state which also has large distributions out of $MoS_2$ layers (Fig. S2b). In $ReS_2$, due to the localized distributions of both the VBM and CBM states within ReS$_2$ layers (Fig. S2c), the band edge positions and bandgaps don't change much with reduced layer numbers, which is in agreement with the experimental observations (*43*). Similarly, the defect states of P$_S$ and Cl$_S$ are also localized within ReS$_2$ layers (Fig. 4a) with nearly unchanged absolute values of the $E_{neu}$, as seen in Fig. 4(d).

In general, the larger delocalization out of 2D material layers a defect state has, the larger change of the defect state towards the band edges with the reduction of dimensionality due to QCE. However, because QCE is always larger on band edges than on defect state, the defect $E_{neu}$ with respect to the band edges always becomes deeper. Nevertheless, we can still conclude that, the more delocalized out of 2D material layers a defect state, the more likely it is to have relatively small defect IE when dimensionality is reduced to monolayer. Consequently, high carrier concentrations in such 2D materials that have defect states with large distributions out of material layers can be more easily realized.

The second reason that causes the larger defect IEs with the reduction of dimensionality is the reduced SE with the decrease of thickness. After the ionization of a neutral defect by accepting one electron from the VBM or donating one electron to the CBM, the electron or hole located at this defect will interact with all the other charges through Coulomb repulsion. In bulk BP, MoS$_2$, and ReS$_2$, the interaction between the charged defects and all the other charges can be screened by the atoms in the whole space. In trilayer, bilayer and monolayer systems, the interaction can only be screened by three, two and one layers of atoms, respectively. The smaller the screening, the more energy it costs for the electronic relaxation. As shown in Fig. 2(e), Fig. 3(e), and Fig. 4(e), our calculations of all the three systems with different strengths of interlayer interaction clearly show that, for all the defects, $E_{ER}$ increases as the number of layers decreases. Our results thus suggest that, to reduce the defect IEs in monolayer 2D semiconductors, increasing the SE will be helpful.

As both $E_{neu}$ with respect to band edges and $E_{ER}$ increase with the reduction of dimensionality due to QCE and SE, respectively, the defect IEs in monolayer systems are much larger than those in few-layer and bulk systems. Note that, from bulk BP to monolayer BP, we find that the lattice constants are slightly stretched (Supplemental

Materials), in agreement with previous calculations (*60*). Consequently, the energy gain due to structural relaxations ($-E_{SR}$) can slightly increase with the dimension reduction (Fig. 2f), thus slightly reducing the defect IEs. However, the increase of such energy gain can hardly dominate over the increase of $E_{neu}$ and $E_{ER}$ brought by the dimension reduction effect in the BP cases. For the MoS$_2$ and ReS$_2$ cases, the structural relaxation energy gain is almost insensitive to layer numbers (Fig. 3f and Fig. 4f) and thus has little effect on the change of defect transition energy levels with reduced dimensionality. Note that, depending on the relative strengths of QCE on band edges and $E_{neu}$ of defects and SE on $E_{ER}$, our results for the above three systems demonstrate our proposed conceptual picture in Fig. 1 that the defect transition energy levels can go either upward or downward in the bandgap with the dimensionality reduction. For example, the acceptor levels of V$_P$ and Si$_P$ in BP go towards the VBM from bulk to monolayer due to the larger changes of $E_{neu}$ caused by QCE, while the donor level of Te$_P$ generally goes away from the CBM due to the larger change of $E_{ER}$ caused by SE. Similarly, all the defect levels in MoS$_2$ and ReS$_2$ go away from the band edges due to the less QCE on $E_{neu}$ than SE on $E_{ER}$.

The QCE and SE not only increase the defect IEs with the decrease of layer thickness, but also affect the transition energy levels of the defects located at different sites when the layer thickness is fixed. As seen in Fig. 2(c), in trilayer BP, the defect transition energy levels with respect to the band edges generally become shallower when the defect locations change from the surface sites to the *in* sites and *in'* sites. For example, the (0/-) level of V$_P$ with respect to the VBM decreases from 0.10 eV at the surface site to 0.08 eV at the *in* site and to 0.03 eV at the *in'* site. Similarly, the (0/-) level of Si$_P$ decreases from 0.16 eV at the surface site to 0.12 eV at the *in* site and to -0.06 eV at the *in'* site. The (0/+) level of Te$_P$ with respect to the CBM decreases from 0.60 eV at the surface site to 0.45 eV at the *in* site and to 0.43 eV at the *in'* site. While the $E_{neu}$ of these defects don't change much (Fig. 2d) because the quantum confinement is little changed, the decrease of the defect IEs from the surface sites to the *in* sites and *in'* sites is mainly dominated by the reduced SE. Apparently, charged defects at the *in'* sites suffer more screening than those at the *in* and surface sites. As a result, the $E_{ER}$ at the *in'* sites is generally smaller than that at the *in* and surface sites (Fig. 2e), leading to the smallest defect IEs at the *in'* sites in trilayer BP. Similar things are also found in MoS$_2$ systems, as shown in Figs. 3(c)-3(d).

**DISCUSSIONS**

Based on the above understanding of QCE and SE on the defect IEs and our calculation results for the three typical systems, strategies to reduce the defect IEs in 2D monolayer semiconductors can be proposed straightforwardly: one is to decrease QCE on the band edges and increase QCE on the defect states; the other one is to increase SE. While QCE, mainly determined by the material intrinsic properties, is hard to influence, some tuning by strain and alloying can still be attempted. In contrast, the SE can be tuned by changing the material dielectric environment.

To demonstrate the feasibility of increasing SE to reduce defect IEs, we adopt the implicit continuum solvation model (*61, 62*) for BP system, which allows us to change the dielectric environments surrounding monolayer BP while keeping the properties of BP monolayer unchanged (Fig. 5a). Our results confirm that the electronic bandgap of BP monolayer doesn't change with the variation of the solvent dielectric constants. We should note that, the dielectric environments can affect exciton binding energies and change material optical bandgaps (*23*). However, the defect IEs, which are referenced to band edges, are generally not affected by exciton effects (*57*). As a result, the exciton effects are neglected in this study (*57*). To focus on screening effect on the defect IEs, we just consider the electrostatic interactions between BP and solvent while omitting the cavitation and dispersion energies. As shown in Fig. 5b, our calculation results clearly show that, with the increase of the solvent dielectric constants $\epsilon_s$ thus the increase of the screening effects, the defect IEs in BP decrease monotonically, similar to the deep-to-shallow level transition of Re and Nb dopants in monolayer MoS$_2$ with dielectric environments (*63*). The decomposition analysis confirms that the reductions of the defect IEs are mainly attributed to the reduced $E_{ER}$ (Fig. 5d) while $E_{neu}$ and $E_{SR}$ keep almost unchanged as expected (Figs. 5c and 5e). Besides, we find that the defect IEs and the $E_{ER}$ are linearly dependent on the $1/\epsilon_s$ (Figs. 5f and 5g), which is expected from the Coulomb interactions. Interestingly, both the defect IEs and $E_{ER}$ will approach some limits when $\epsilon_s$ approaches the infinity, which is determined by the finite in-plane screening due to BP itself. With the increased SE due to the continuum solvent and thus the decrease of defect IEs, charge carrier concentration is expected to be improved. For example, for intrinsic BP monolayer with $V_P$ (0/-) level of 0.29 eV at $\epsilon_s = 10$, the hole density can reach $9.55 \times 10^7\ cm^{-2}$ at T = 300 K, more than 5 orders of magnitudes larger than the case when BP is placed at vacuum.

In reality, the dielectric environment can be provided by either substrate or by encapsulating layers. We note that the heterostructure of hexagonal boron nitride ($h$—BN) monolayer encapsulated phosphorene, like BN/BP/BN, have been experimentally synthesized (64), providing us a practical example to study the SE on the defect IEs. The structural model of BN/BP/BN heterostructure is shown in Fig. 6(a), where lattice constants are fixed to be the same as monolayer BP and $h$—BN supercells are applied by biaxial strains of less than 7% to match a $3 \times 1$ supercell of monolayer BP (see atomic positions in the Supplemental Materials). In this case, the heterostructure band edges are still mainly derived from the BP layer due to the wide bandgap of $h$—BN (see Fig. 5). However, due to the Van der Waals interaction between BP and $h$—BN, both the VBM and CBM of the heterostructure are pushed upward compared to the band edges of bare monolayer BP, as seen in Fig. 7. This is because the $p_z$ orbitals of nitrogen atoms lie closely below the $p_z$ orbitals of phosphorus atoms (see Figs. 6c and 6e). The extra coupling between nitrogen $p_z$ orbitals and phosphorus $p_z$ orbitals thus lift up both the VBM and CBM of the heterostructure with the occupied VBM state increased by 0.03 eV and the unoccupied CBM state increased by about 0.24 eV.

Despite an increased bandgap in BN/BP/BN heterostructure, our calculations clearly show that the defect transition energy levels in $h$—BN encapsulated BP become shallower compared to those in bare BP, in agreement with our conceptual picture and the above continuum solvent model simulations. As shown in Fig. 7(a), the (0/-) levels of $V_P$ and $Si_P$ with respect to the VBM decrease from 0.57 eV and 0.68 eV in bare monolayer BP to 0.43 eV and 0.31 eV in the heterostructure, respectively. The (0/+) levels of $Te_P$ with respect to the CBM decrease from 0.75 eV in bare monolayer BP to 0.58 eV in the heterostructure. The reduction of the defect transition energy levels from bare monolayer BP to the BN/BP/BN heterostructure is mainly dominated by the increased SE in the heterostructure as expected, which is demonstrated by the largely reduced electronic relaxation energies. As seen in Fig. 7(b), the neutral single-electron point defect levels exhibit very small change, especially for the acceptors, which is expected as QCE doesn't change. Note that, for the donor Te$_P$, the relatively large change of $E_{neu}$ is induced by the large decrease of the vertical Te─P bond (2.835 Å in the heterostructure versus 3.045 Å in the bare BP). Similar to the case when BP changes from few-layer to bulk (see Fig. 2d and the above discussions), the coupling between coupling of Te $p_z$ orbitals and P $p_z$ orbitals through vertical Te─P gets much stronger, leading to the decreased neutral single-electron defect levels of Te$_P$ with respect to the CBM. In contrast to $E_{neu}$, $E_{ER}$'s show large drops for all the three defects, as clearly seen in Fig. 7(c), which result from the increasing SE due to the encapsulating $h$─BN layers. Meanwhile, the structural relaxation energy gains don't change much as expected (Fig. 7d). Note that, due to the computation limit, we only consider encapsulation of BP by one $h$─BN layer at each side. In this case, by assuming the Fermi level is pinned at the V$_P$ (0/-) level of 0.43 eV, the intrinsic BP monolayer can reach a hole density of $4.38 \times 10^5$ $cm^{-2}$ at T = 300 K, which is more than 2 orders of magnitudes larger than the bare BP layer. We expect the defect transition energy levels in monolayer BP can be further reduced and the carrier densities can be further enhanced by increasing the screening using more encapsulating $h$─BN layers.

In conclusion, we develop a conceptual picture of dimensional effects on the defect IEs and demonstrate it using first-principles calculations of BP, MoS$_2$, and ReS$_2$ systems. While the quantum confinement makes the neutral single-electron point defect levels deep, the reduced screening leads to high energy cost for the electronic relaxation. Based on the gained insight into defect behaviors, different strategies can be explored in achieving more efficient carrier doping in 2D semiconductors. While QCE, mainly determined by the material intrinsic properties, is hard to influence, some tuning by strain and alloying can be attempted. In contrast, the SE can be controlled by dielectric environment. By embedding BP monolayer into continuum solvent or encapsulating it between two *h*─BN layers, as practical examples, we demonstrate the feasibility of increasing the screening to reduce the defect IEs and thus enhance the charge carrier doping efficiency. Our simulations show that, the hole density in intrinsic phophorene can be enhanced by 5 orders of magnitudes when BP monolayer is placed in a continuum solvent with a dielectric constant of 10 or by 2 orders of magnitudes when BP monolayer is encapsulated by two *h*─BN layers. The carrier concentration is expected to be further improved by considering better dopants and dielectric environment with larger screening effects.

**MATERIALS AND METHODS**

Our first-principles total energy calculations are performed using density-functional theory (DFT) (*65, 66*) as implemented in the VASP code (*67, 68*). The bandgaps of BP systems are corrected by Heyd-Scuseria-Ernzerhof (HSE06) hybrid functional (*69*), which yields bandgaps of 0.30 eV, 0.78 eV, 1.05 eV and 1.60 eV for bulk, trilayer, bilayer and monolayer BP, respectively, in good agreement with recent experimental results (*70*). For $MoS_2$ and $ReS_2$, we simply adopt the generalized gradient approximation (GGA) formulated by Perdew, Burke, and Ernzerhof (PBE) (*71*), which gives bandgap values of 1.68 eV, 1.11 eV, 1.45 eV, and 1.29 eV for monolayer $MoS_2$, bulk $MoS_2$, monolayer $ReS_2$, and bulk $ReS_2$, respectively, in agreement with experimental works (*43,49*). For defect calculations in charged 2D systems, we adopt the method proposed by Wang *et al.* (*21*) to get the converged total energies (see also Supplemental Materials) ,which has been proved to perform well for 2D charged systems and achieve good agreement with the method proposed by Komsa *et al.*(*72, 73*). The band edge levels and band alignments are obtained by setting the vacuum level as zero. Other detailed calculation parameters are given in the Supplemental Materials.

**SUPPLEMENTARY MATERIALS**

Supplementary materials for this article are available at XXXX.

Detailed calculation methods and convergence test, detailed calculated data of structural information and level values, and other supporting data.

Fig. S1. Band alignments from monolayer systems to few-layers and bulk systems for BP, $MoS_2$, and $ReS_2$.

Fig. S2. Partial charge densities of band edges of monolayer BP, $MoS_2$, and $ReS_2$.

Fig. S3. Convergence test of calculated defect transition energy levels.

Table S1. Detailed data related to the defect transition energy levels of BP systems.

Table S2. Detailed data related to the defect transition energy levels of $MoS_2$ systems.

Table S3. Detailed data related to the defect transition energy levels of $ReS_2$ systems.

**Figures**

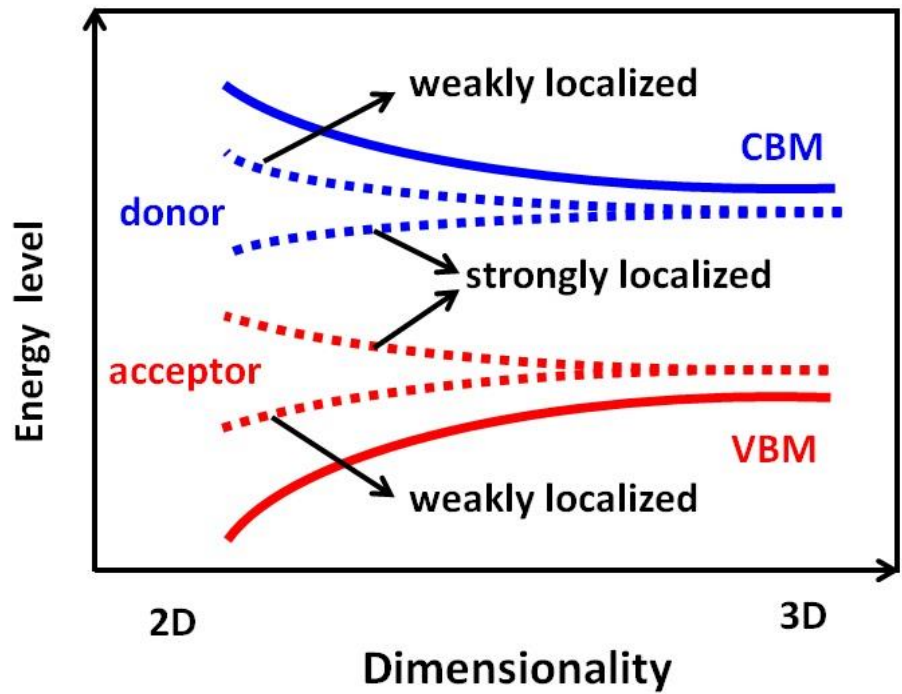

**Fig. 1. A conceptual diagram to show how band edges and defect transition energy levels with different localization characters change with the dimensionality of semiconductor systems.**

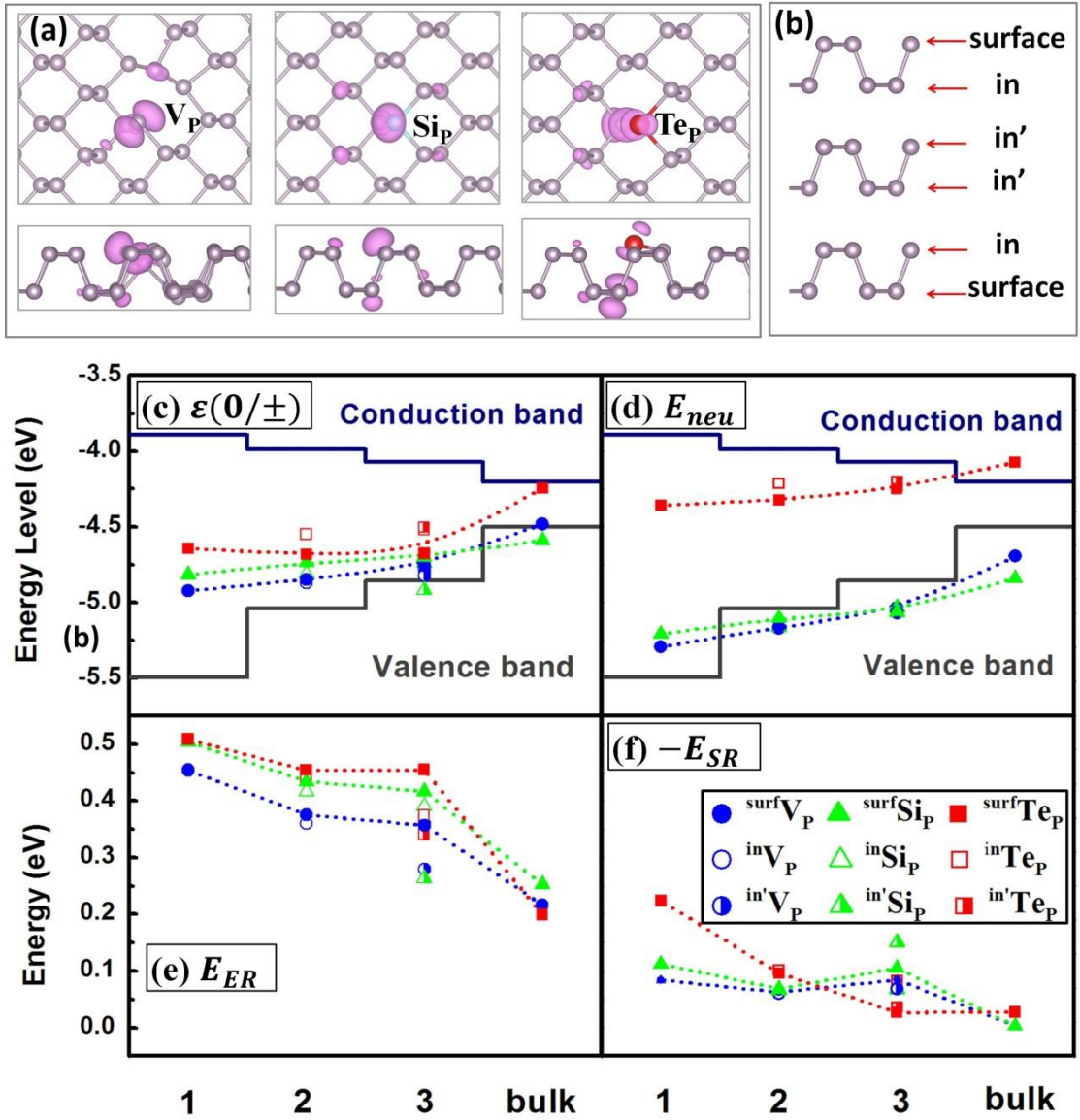

**Fig. 2. Structure models and defect properties of bulk, mono- and few-layer BP.** (a) Structures and partial charge densities of $V_P$, $Si_P$, and $Te_P$. P, Si, and Te atoms are colored in light purple, cyan, red, respectively. The defect partial charge densities are shown in pink with an isosurface of 0.005e. (b) Structure of few-layer BP with different possible defect positions labled as surface, *in* and *in'*. (c) Defect transition energy levels, (d) neutral single-electron point defect levels, (e) electronic relaxation energy cost and (f) structural relaxation energy gain of point defects at different defect positions in mono-, bi-, tri-layer and bulk BP. The levels are given with respect to the vacuum levels. The defect levels and relaxation energies of surface defects and the bulk defects are connected by dashed lines for guidance.

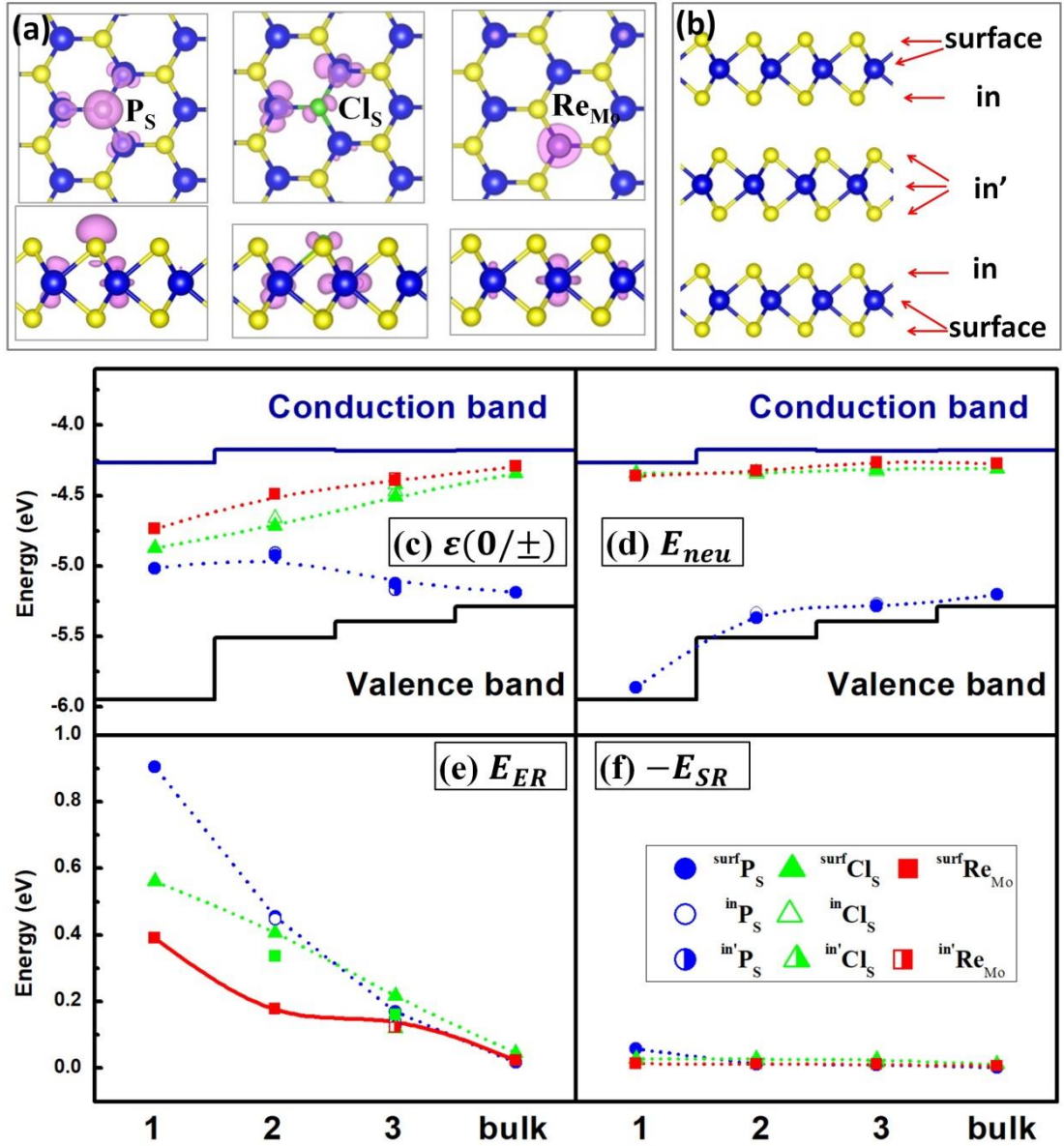

**Fig. 3. Structure models and defect properties of bulk, mono- and few-layer MoS$_2$.** (a) Structures and partial charge densities of P$_S$, Cl$_P$, and Re$_{Mo}$. Mo, S, P, Cl, and Re atoms are colored in blue, yellow, light purple, green, and dark purple, respectively. The defect partial charge densities are shown in pink with an isosurface of 0.007e. (b) Structure of few-layer MoS$_2$ with different possible defect positions labled as surface, *in* and *in'*. (c) Defect transition energy levels, (d) neutral single-electron point defect levels, (e) electronic relaxation energy cost and (f) structural relaxation energy gain of point defects at different defect positions in mono-, bi-, tri-layer and bulk MoS$_2$. The levels are given with respect to the vacuum levels. The defect levels and relaxation energies of surface defects and the bulk defects are connected by dashed lines for guidance.

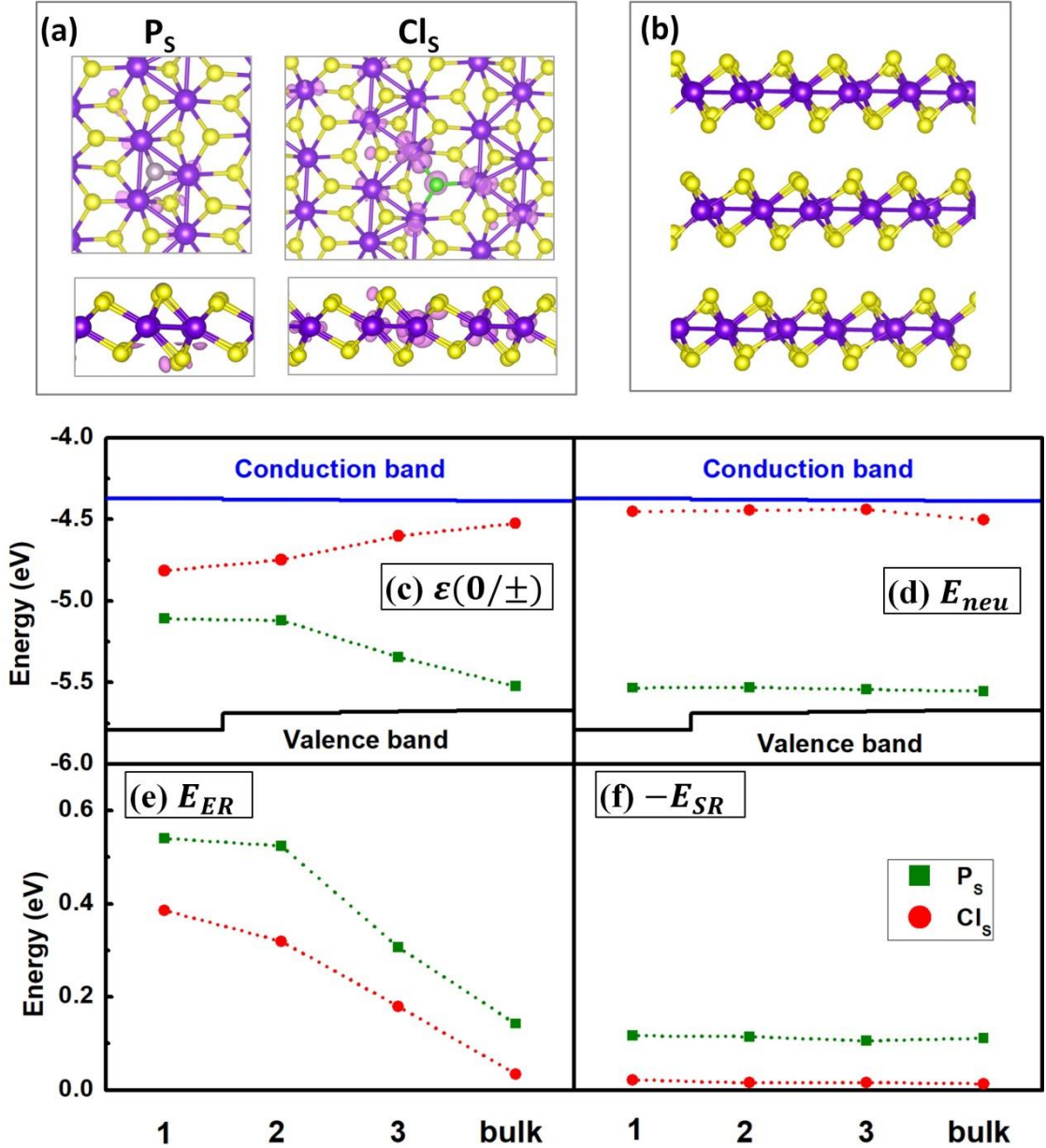

**Fig. 4. Structure models and defect properties of bulk, mono- and few-layer ReS$_2$.** (a) Structures and defect partial charge densities of P$_S$ and Cl$_S$. Yellow, dark purple, light purple, and green balls are S, Re, P, and Cl atoms, respectively. The defect partial charge densities are shown in pink with an isosurface of 0.003e. (b) Structure of mono-, bi-, and tri-layer ReS$_2$ used in this work. (c) Defect transition energy levels, (d) neutral single-electron point defect levels, (e) electronic relaxation energy cost and (f) structural relaxation energy gain of point defects at different defect positions in mono-, bi-, tri-layer and bulk ReS$_2$. The levels are given with respect to the vacuum levels. The defect levels and relaxation energies of surface defects and the bulk defects are connected by dashed lines for guidance.

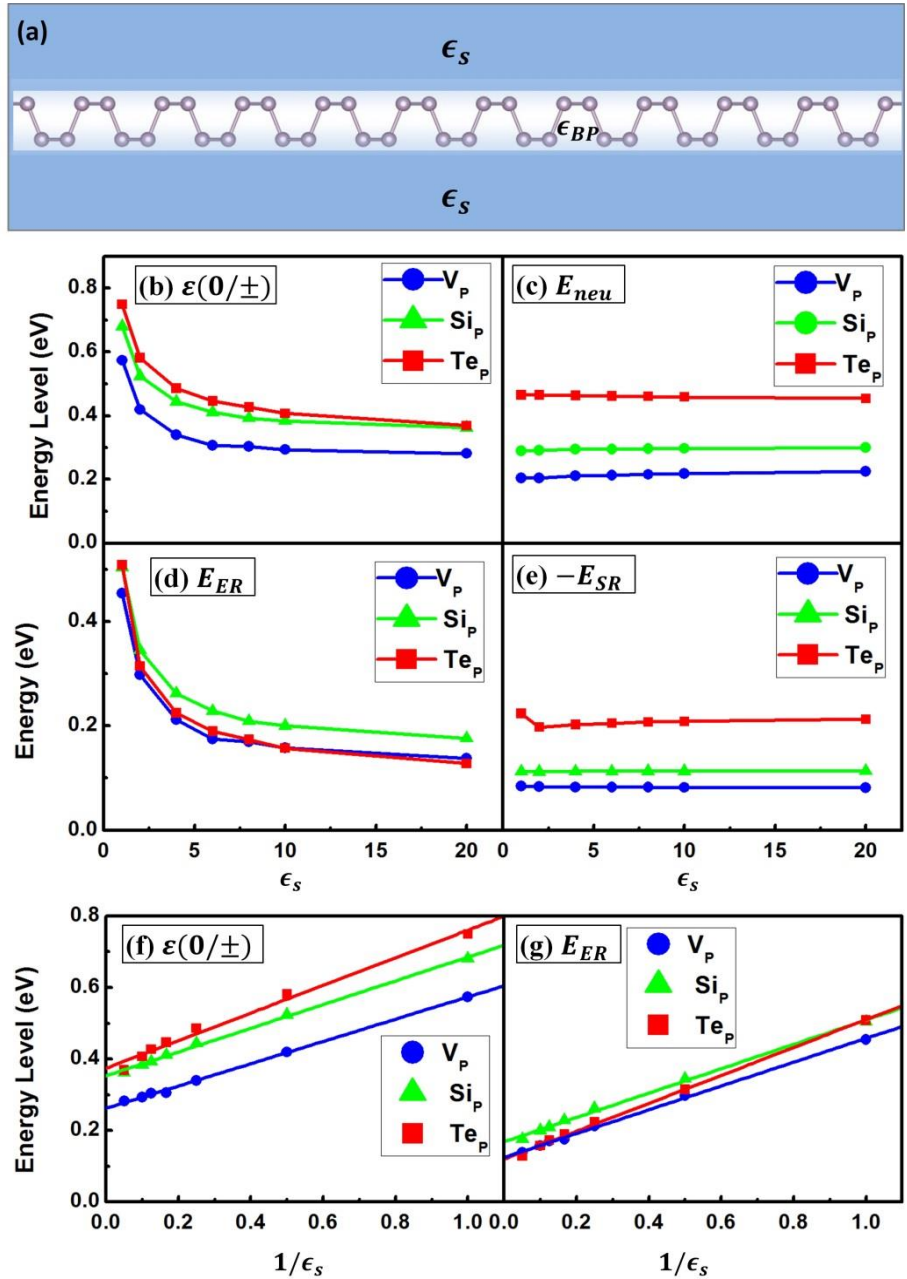

**Fig. 5. Continuum solvation model and defect properties of monolayer BP in a dielectric environment.** (a) Solvation model of BP in a dielectric environment. P atoms are colored in light purple and the light blue color area stands for solution with a dielectric constant of $\epsilon_s$. (b) Defect transition energy levels, (c) neutral single-electron point defect levels, (d) electronic relaxation energy cost and (e) structural relaxation energy gain of point defects in BP as functions of the dielectric constants of the solvent. (f) Defect ionization energies and (g) $E_{ER}$ linear dependence on the $1/\epsilon_s$.

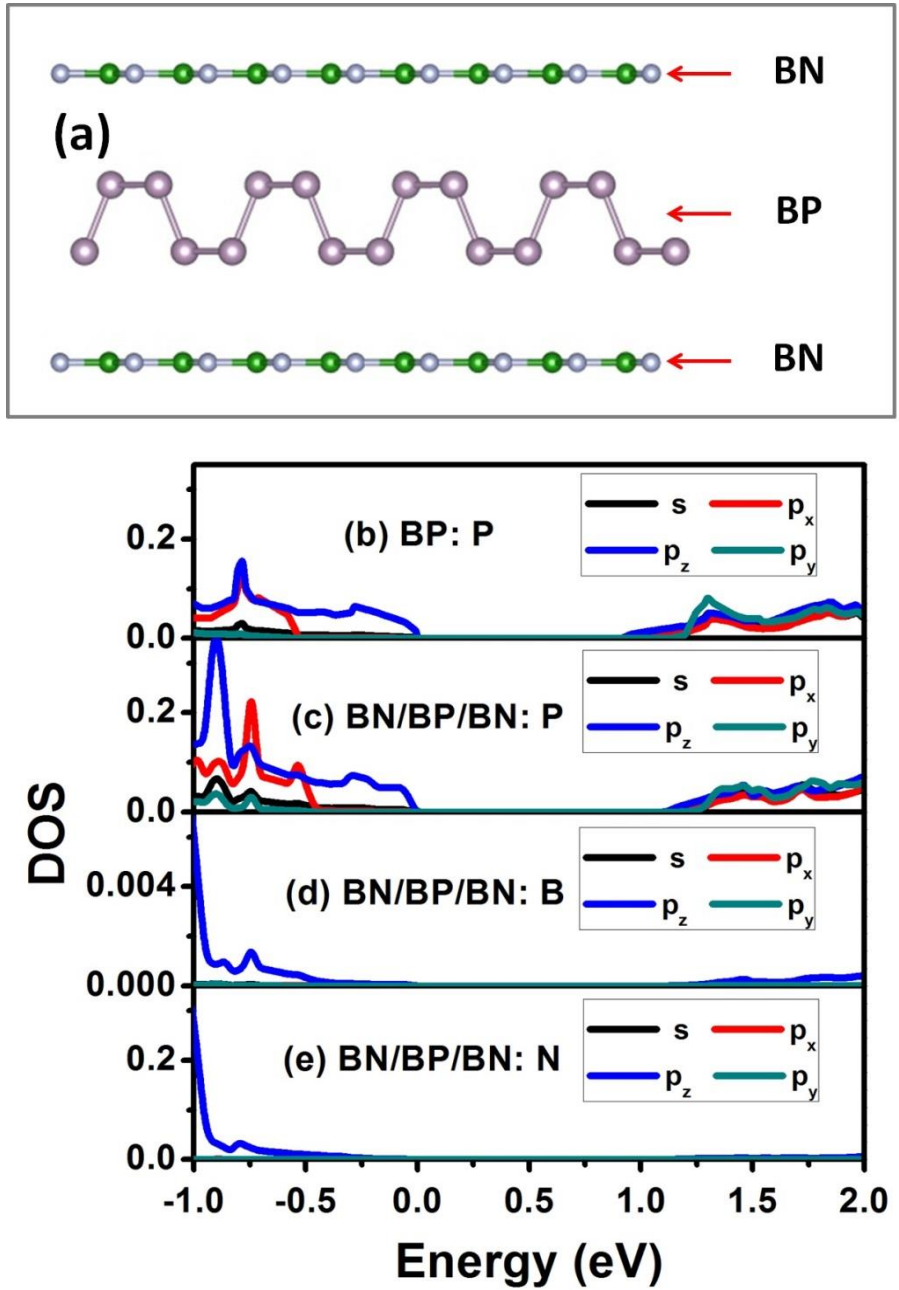

**Fig. 6. Structure model and density of states of BN/BP/BN heterostructure**. (a) Structural model of BN/BP/BN heterostructure. P atoms are colored in light purple, B atoms are green, and N atoms are light blue. Atomic projected partial density of states for (b) P atoms in BP, (c) P atoms in BN/BP/BN, (d) B atoms in BN/BP/BN, and (e) N atoms in BN/BP/BN. The VBM states are set as zero.

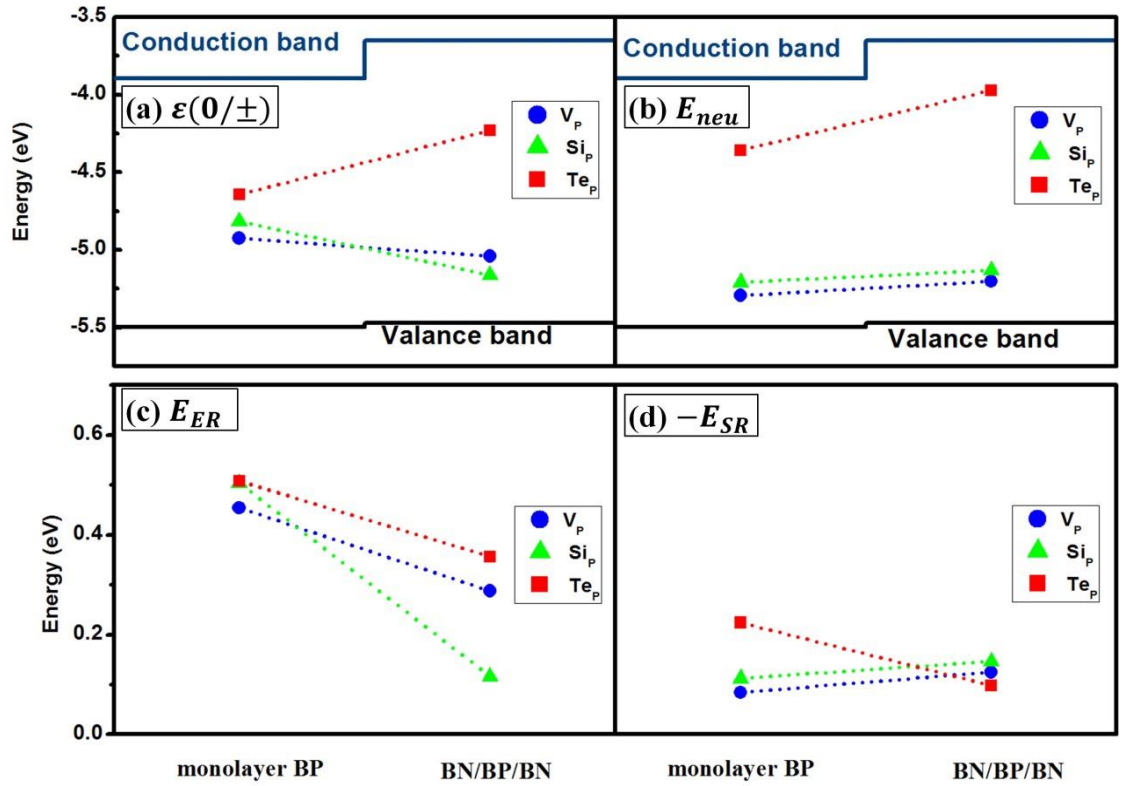

**Fig. 7. Defect properties of monolayer BP encapsulated by BN layers.** (a) Defect transition energy levels, (b) neutral single-electron point defect levels and (c) electronic relaxation energies of point defects positions in monolayer BP and BN/BP/BN heterostructure. The defect levels and electronic relaxation energies of defects are connected by dashed lines for guidance. All the levels are given with respect to the vacuum levels.